\documentclass[aps,twocolumn]{revtex4}

\usepackage[dvipdfm]{graphicx}
\usepackage{subfigure}
\usepackage{dcolumn}

\begin{document}


\title{A Q-Ising model application for linear-time image segmentation}

\author{Frank W. Bentrem}
\email{frank.bentrem@nrlssc.navy.mil}
\affiliation{Center for Bio/Molecular Science and Engineering,~Naval Research Laboratory, Washington, DC 20375 \\ Department of Physics and Engineering Physics, Tulane University, New Orleans, Louisiana 70118 \\ Marine Geosciences Division, Naval Research Laboratory, Stennis Space Center, Mississippi 39529}
\thanks{Present address}


\date{\today}

\begin{abstract}
A computational method is presented which efficiently segments digital grayscale images by directly applying the Q-state Ising (or Potts) model. Since the Potts model was first proposed in 1952, physicists have studied lattice models to gain deep insights into magnetism and other disordered systems. For some time, researchers have realized that digital images may be modeled in much the same way as these physical systems (i.e., as a square lattice of numerical values). A major drawback in using Potts model methods for image segmentation is that, with conventional methods, it processes in exponential time. Advances have been made via certain approximations to reduce the segmentation process to power-law time. However, in many applications (such as for sonar imagery), real-time processing requires much greater efficiency. This article contains a description of an energy minimization technique that applies four Potts (Q-Ising) models directly to the image and processes in linear time. The result is analogous to partitioning the system into regions of four classes of magnetism. This direct Potts segmentation technique is demonstrated on photographic, medical, and acoustic images.
\end{abstract}


\maketitle


\section{Introduction}
Segmenting digital images into regions of distinct types has applications in a great many fields, e.g. medical imaging \cite{pham00,yang02,peng03,grau04}, surveillance by synthetic aperture radar \cite{descombes96} and satellite, and underwater acoustic imaging (.e.g. see Ref.~\onlinecite{bentrem06}), to name a few. In general, image segmentation is performed by classifying image regions by color, intensity, and texture. Of course, only the latter two are considered in grayscale segmentation, which is the subject we consider in this article. While classification by intensity is a straightforward assessment of the brightness/darkness of an image pixel or group of pixels (as with histogram segmentation methods \cite{pham00}), texture classification is much more complex \cite{pham00,asano01}. Although it is difficult to define image texture precisely, we can say that it is the spatial relationship of the intensities (i.e. ``graininess'') of an image region. Clearly, there are indeed a large number of possible texture types in digital images. 

\begin{figure}
\begin{center}
\includegraphics*[width=3.375in]{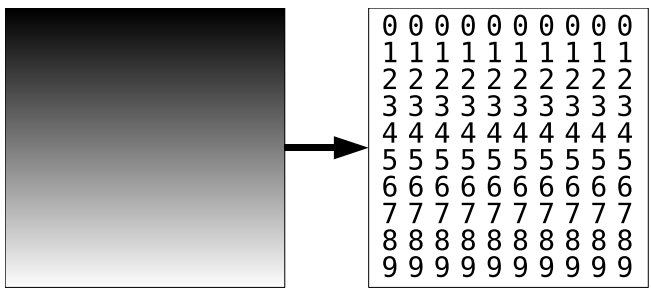}
\end{center}
\caption{\label{fig:gradient} On left, a grayscale image varies from black to white in going from top to bottom. The image is represented as a matrix of intensity values as illustrated on the right.}
\end{figure}

A grayscale digital image may be represented as a matrix of numerical values, as in Fig.~\ref{fig:gradient}, which indicate the intensity or brightness (gray level) of the corresponding image pixel. Techniques for image segmentation (such as thresholding and histogram methods) that focus solely on intensity are the most computationally efficient since they generally require just one or two passes through the intensity matrix. Identifying regions of different texture, however, tend to be relatively expensive in terms of computation time \cite{asano01}. Here I describe four types of magnetism using the Ising model \cite{ising25} and show how its generalization, the Q-Ising (or Potts) model \cite{potts52}, can be used analogously to segment grayscale images into four categories \cite{bentrem09}. The common approach (e.g. see \cite{peng03,descombes96,tanaka02,neirotti03}) to segmentation using the Potts model applies the Potts model to the categorizing labels, i.e., the segmented image. (A similar approach is also used for more general data clustering \cite{blattWD,wiseman98}.) The optimal segmentation is often found using a Monte Carlo approach, which processes in exponential time (time for computer processing increases exponentially with the number of pixels contained in the digital image \cite{tanaka02}). Use of the Bethe approximation \cite{tanaka02,tanaka04} allows an optimal segmentation to be obtained in power-law time. However, the direct Potts segmentation method presented in this article, which utilizes both intensity and texture, processes in \textit{linear time}--among the fastest methods currently used. (Other unrelated linear-time or nearly linear-time segmentation methods are described in \cite{sharon00,felzenszwalb04,falcao06}, though, these contain undesirable feature for many applications. For example, graph partitioning methods \cite{felzenszwalb04,falcao06} cannot relate disjoint sets with equivalent textures and require training sets.) This computational speedup arises from applying the Potts model directly to the \textit{original image itself} rather than to the segmented representation. The speedup is important for high-data-volume/high-resolution applications where time is constrained.

\section{Ising Model}

In 1925, Ernst Ising proposed the model \cite{ising25,chandler87,lee04}, which now bears his name, as a means to study phase transitions in magnetism. Ising solved the original Ising model (only two states--namely 1, or spin up, and -1, or spin down) in one dimension. However, the one-dimensional Ising model did not exhibit a phase transition, and the two-dimensional Ising model was not solved until 1944. At that time Onsager \cite{onsager44} demonstrated that the two-dimensional Ising model did, in fact, exhibit an order/disorder phase transition. The Ising model and related models have been widely used to study magnetism and other order/disorder phenomena in physical systems. The two-dimensional model is described here, and it will be used to define four classes of magnetism. The image texture analogs will be presented in a subsequent section.

\begin{figure}
\begin{center}
\includegraphics*[width=3.375in]{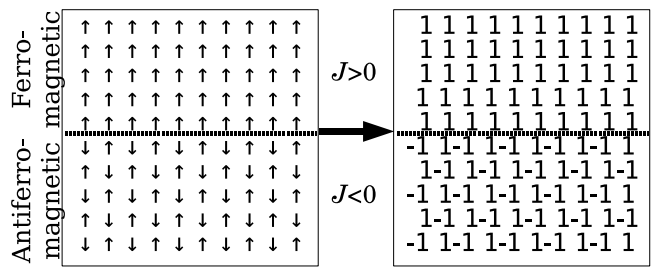}
\end{center}
\caption{\label{fig:ising} A spin matrix is shown on the left with arrows indicating elements whose magnetic states are either spin-up and spin-down. The matrix on the right represents the same matrix with the arrows exchanged for 1's and (-1)'s. In the absence of an external magnetic field ($H=0$), a ferromagnet's spins will spontaneously align as depicted in the top half of the matrices, while the spins for antiferromagnets will oppositely align (alternate in opposite directions) as depicted in the bottom half of the matrices.} 
\end{figure}

\begin{figure}
\begin{center}
\includegraphics*[width=3.375in]{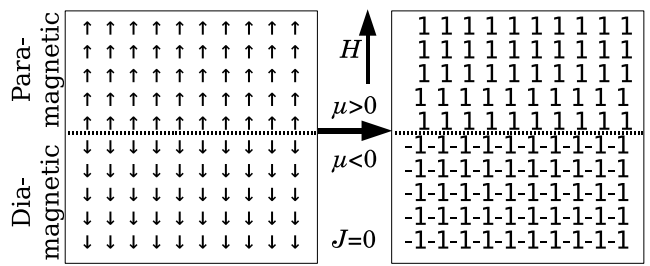}
\end{center}
\caption{\label{fig:ising2} Magnetic spins are aligned with (left, top) and against (left, bottom) the magnetic field $H$. These regions are associated with paramagnetism and diamagnetism, respectively.}
\end{figure}

In the Ising model, given a two-dimensional lattice of magnetic spins oriented either up or down as in Fig.~\ref{fig:ising} (numerically represented as a matrix of 1's and $-1$'s), the total energy $E$ associated with the lattice is

\begin{equation}
\label{eq:ising}
E = -\mu H \sum_i s_i - J \sum_{ij, \text{ nn}} s_is_j,
\end{equation}

\noindent
where $i$ represents a magnetic element, $s_i$ is the value of the magnetic spin of $i$ (1 or $-1$), $\mu$ is the magnetic moment of the element, $H$ is the external magnetic field, and $J$ is the interaction coupling constant. The sum over all nearest-neighbors pairs $i$,$j$ is represented by $\sum_{ij, \text{ nn}}$. Nearest neighbors are considered to be the elements directly above, below, left, and right of a given element. If no coupling interaction is present ($J=0$) then $\mu>0$ creates an energetically favorable condition for the spins to align in the direction of the external field $H$, whereas $\mu<0$ leads to spins aligned in the direction opposite of $H$. (See Fig.~\ref{fig:ising2}.)

\section{Four Classes of Magnetism}

In common language, magnetism usually refers to the behavior of magnets that attract iron (known as ferromagnets). Other materials, however, reveal different magnetic behavior. These behaviors may be classified into the following four groups:

\begin{enumerate}

\item{\textit{Ferromagnetism} is exhibited by materials for which, even in the absence of an external magnetic field, a net magnetization is present with north and south poles. A ferromagnet will both attract and repel another ferromagnet depending on the alignment of their poles.}
\item{\textit{Antiferromagnetism} is displayed by materials which do not form magnetic poles even with a modest external magnetic field present. An antiferromagnet will neither attract nor repel another antiferromagnet. Likewise an antiferromagnet will have no magnetic interaction with a ferromagnet.}
\item{\textit{Paramagnetism} shows no spin ordering in the absence of an external field, although the paramagnet's spins align in the same direction of an external field when present. A paramagnet will neither attract nor repel another paramagnet, but will be attracted to a ferromagnet}
\item{\textit{Diamagnetism} also shows no ordering in the absence of an external field; however, the diamagnet's spins will align in the \textit{opposite} direction of any external field. A diamagnet will neither attract nor repel another diamagnet and will be repelled by a ferromagnet.}

\end{enumerate}

\section{Q-Ising Model}
\begin{figure}
\begin{center}
\subfigure[]{
\includegraphics*[width=1.597in]{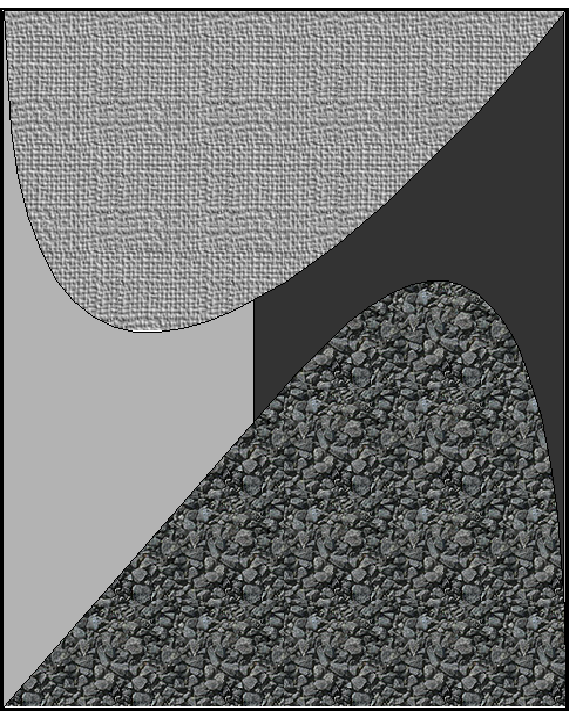}
}
\subfigure[]{
\includegraphics*[width=1.597in]{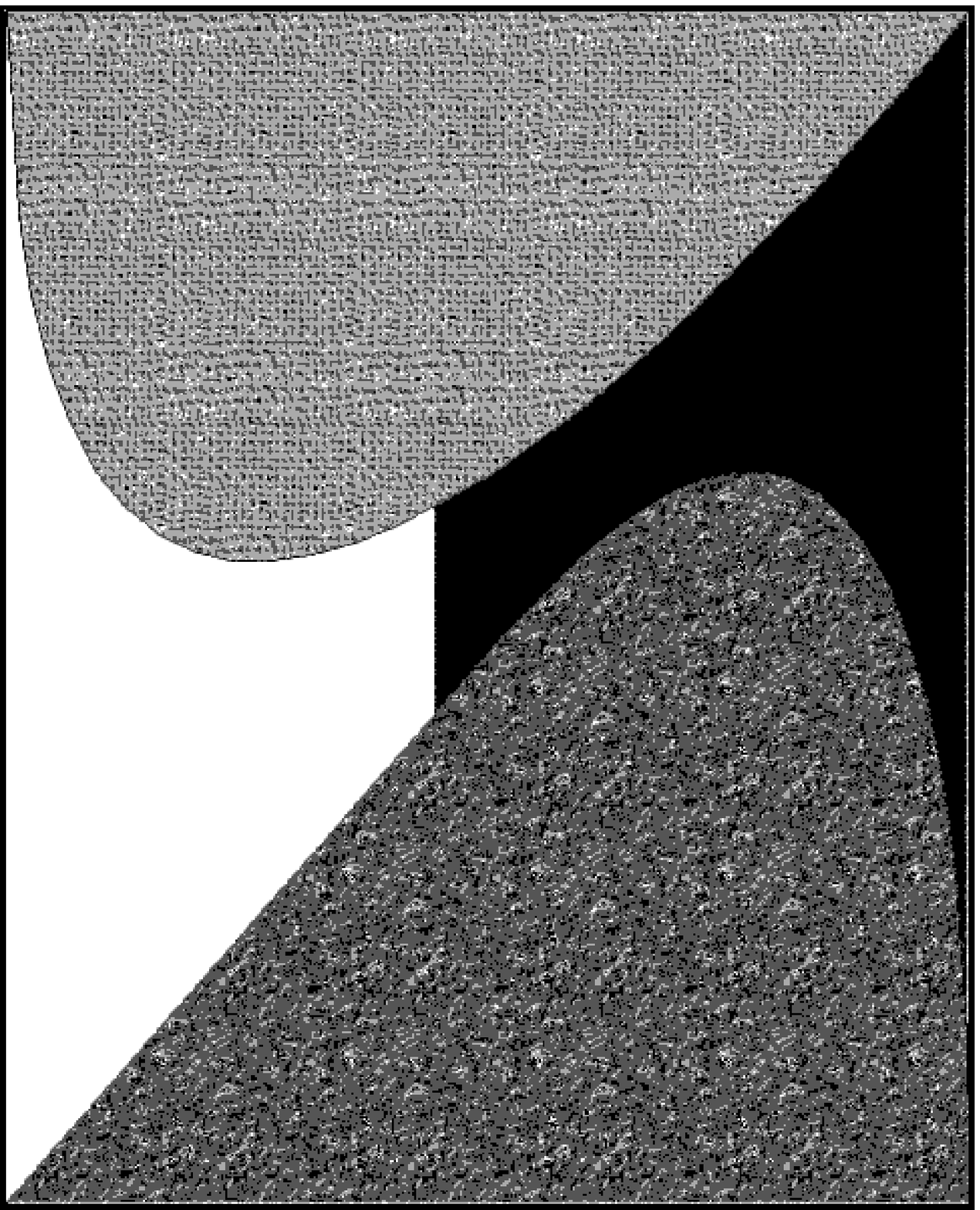}
}
\\
\subfigure[]{
\includegraphics*[width=1.597in]{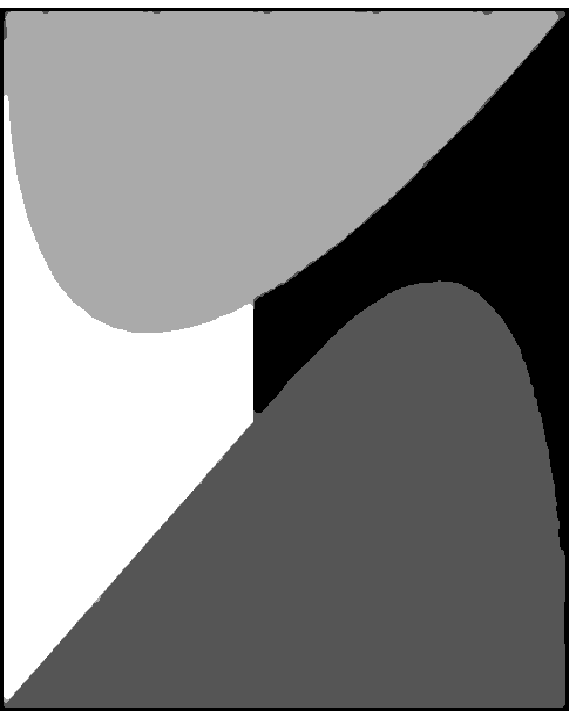}
}
\subfigure[]{
\includegraphics*[width=1.597in]{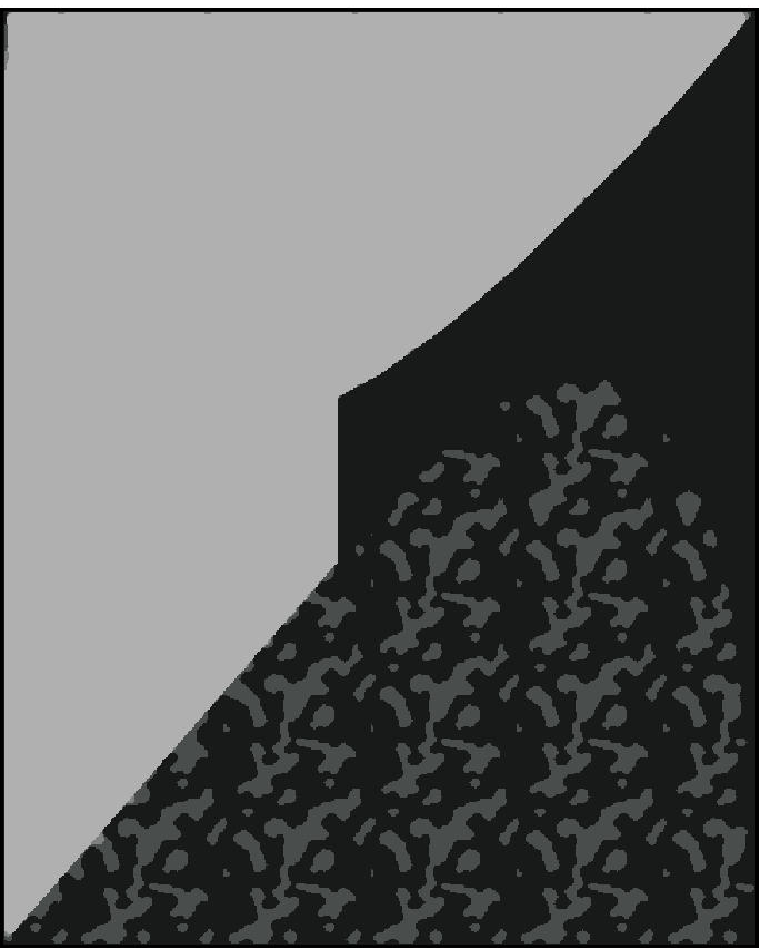}
}
\end{center}
\caption{\label{fig:segment}Sample grayscale image (a) with regions of visibly distinct intensity and texture. Upper right (b) is the segmented image using the direct Potts method with $Q=16$ and $s_t=7.5$. At lower left (c), the segmented image from (a) is filtered with a despeckle algorithm to reveal four distinct texture categories. The categories are bright and smooth (white), bright and grainy (light gray), dark and grainy (dark gray), and dark and smooth (black). Lower right (d) is the despeckled segmentation result produced using the four-level histogram approach. The histogram method does not properly distinguish regions of different textures, which demonstrates the problem in using the histogram approach for texture analysis.}
\end{figure}

This section describes the Q-Ising model, also known as the Potts model, which will be used to analyze grayscale images where the pixel intensities are considered in place of the magnetic ``spin'' values. The Potts model generalizes the (2-state) Ising model into a $Q$-state model by replacing the 1's and -1's with the numbers $0,1,2,3,\dots,Q-1$ (the pixel intensities in a grayscale image). According to this model, the total energy associated with the system is

\begin{equation}
\label{eq:potts}
E = -\mu H \sum_i (s_i-s_t) - J \sum_{ij, \text{ nn}} \delta(s_i,s_j)
\end{equation}

\noindent
where $s_t$ is the spin threshold (separating light from dark with $0<s_t<Q-1$) and the kronecker delta function $\delta$ (1 if the elements are equal, 0 otherwise) replaces the summand for the second summation in Eq.~\ref{eq:ising}. Now lets suppose that a magnetic spin matrix represents the intensities in a grayscale image and that a uniform external magnetic field, $H=1$, is applied to the system. We wish to find a set of pairs $(\mu_\alpha ,J_\alpha)$ for Eq.~\ref{eq:potts} which characterizes the regions $\alpha$ in the spin matrix. Then each spin in the matrix will be associated with a particular parameter pair. The ``optimal characterization'' is taken to be the partitioning of the spin matrix in such a way that the total energy of the matrix is minimized. Minimizing Eq.~\ref{eq:potts} requires the set of four parameter pairs $(\mu>0,J>0)$, $(\mu>0,J<0)$, $(\mu<0,J>0)$, and $(\mu<0,J<0)$. For convenience, and without loss of generality, we simply use $(\mu_1=1,J_1=1)$, $(\mu_2=1,J_2=-1)$, $(\mu_3=-1,J_3=1)$, and $(\mu_4=-1,J_4=-1)$ for what we will call regions $\alpha=1,2,3,4$ respectively. The energies for the regions can be written as

\begin{equation}
E_1 = -\sum_{i\in A1} (s_i-s_t) - \sum_{{i\in A1}, j\in\text{nn}} \delta(s_i,s_j),
\label{eq:alpha1}
\end{equation}
\begin{equation}
E_2 = -\sum_{i\in A2} (s_i-s_t) + \sum_{{i\in A2}, j\in\text{nn}} \delta(s_i,s_j), 
\label{eq:alpha2}
\end{equation}
\begin{equation}
E_3 = \sum_{i\in A3} (s_i-s_t) - \sum_{{i\in A3}, j\in\text{nn}} \delta(s_i,s_j),
\label{eq:alpha3}
\end{equation}
\begin{equation}
E_4 = \sum_{i\in A4} (s_i-s_t) + \sum_{{i\in A4}, j\in\text{nn}} \delta(s_i,s_j),   
\label{eq:alpha4}
\end{equation}

\noindent
where $A\alpha$ is the set of all spins $i$ labeled with category $\alpha$. The $j$'s in the interaction terms are nearest-neighbor spins that may or may not belong to $A\alpha$. Then the total energy $E$ is simply the sum over all regions,

\begin{equation}
E = \sum_\alpha E_\alpha,
\label{eq:total-energy}
\end{equation}

\noindent
where $\alpha = 1,2,3,4$ are the region segments assigned to categories 1, 2, 3, and 4, respectively. These energies correspond to regions which are 1) paramagnetic and ferromagnetic, 2) paramagnetic and antiferromagnetic, 3) diamagnetic and ferromagnetic, and 4) diamagnetic and antiferromagnetic. Minimizing the total energy (Eq.~\ref{eq:total-energy}) effectively partitions the spin matrix into regions corresponding to these four classes of magnetism. Reverting back to the grayscale image, these categories correspond to image regions which are 1) bright and smooth, 2) bright and ``grainy'', 3) dark and smooth, and 4) dark and ``grainy''. The direct Potts segmentation process requires the matrix to be separated into regions of these four types in such a way as to minimize the total energy. An example image with four distinct intensity and texture regions is depicted in Fig.~\ref{fig:segment}a and the result from the direct Potts segmentation is shown in Fig.~\ref{fig:segment}b. The parameters used in this example are $Q=16$ and $s_t=7.5$, and the segmented image is shown with four shades of gray representing the four texture categories. The segmented image is filtered using the ImageMagick\footnote[3]{ImageMagick Studio LLC, Landenberg, PA, USA} despeckle algorithm and is shown in Fig.~\ref{fig:segment}c. For comparison, a segmented image using a simple four-level histogram method is shown in Fig.~\ref{fig:segment}d. The four-level histogram method performs poorly in distinguishing the image texture regions.

\section{Algorithm}

To segment a digital grayscale image, we first transform the image into a matrix of intensity values {$s_{i,j}$} as illustrated in Fig.~\ref{fig:gradient}. Since it is common for 8-bit digital grayscale images to contain 256 gray levels ($0 \leq s_{i,j} \leq 255$), reducing the gray levels to $16 \leq Q \leq 64$ usually helps in identifying visual textures. Next we must specify an intensity threshold $s_t$, which separates bright pixels from dark pixels. For each of the examples in this article, the threshold is simply $s_t=(Q-1)/2$, or halfway between maximum and minimum intensity. (If texture identification at a certain scale is desired, one may rescale the image.) Then to perform the energy minimization described above, we follow these two steps for each of the image pixels ($i,j$):

\begin{enumerate}

\item{If $s_{i,j} > s_t$ then ($i,j$) is assigned either texture category 1 or category 2. Otherwise, ($i,j$) is assigned texture category 3 or 4.}

\item{Check the nearest-neighbor values (i.e., the pixels directly above, below, left, and right) of pixel ($i,j$). If one or more nearest-neighbor values are equal to $s_{i,j}$, then pixel ($i,j$) is assigned texture category 1 or 3. Otherwise, ($i,j$) is assigned either texture category 2 or category 4).}  

\end{enumerate}

\noindent
This algorithm guarentees that the total energy defined by Eq.~\ref{eq:total-energy} is minimized and processes in linear time since each pixel requires only a fixed number of computations. Only a single comparison and assignment required in Step 1, and a maximum of four comparisons (nearest neighbors) and assignment required in Step 2. Note that since the energy associated with each pixel is independent of the labeling of its neighbors, choosing an energy minimum for every pixel guarantees a total energy minimum (Eq.~\ref{eq:total-energy}) for the entire image. Finally, the segmented image is smoothed by applying the ImageMagick despeckle algorithm. This serves to eliminate much of the statistical noise in the image.

\section{Results}

\begin{figure}
\begin{center}
\subfigure[]{
\includegraphics[width=1.597in,height=6.0cm]{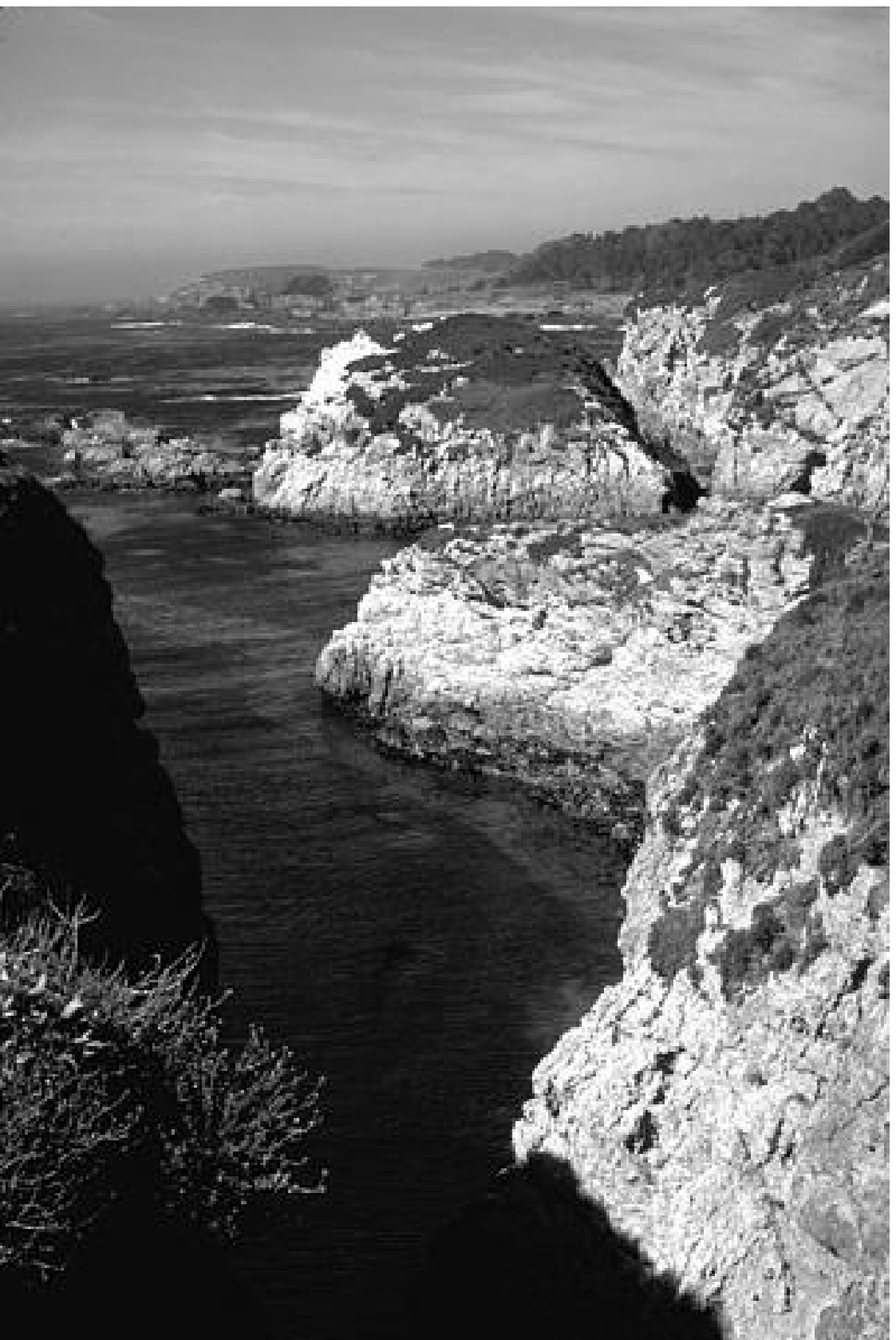} 
}
\subfigure[]{
\includegraphics[width=1.597in]{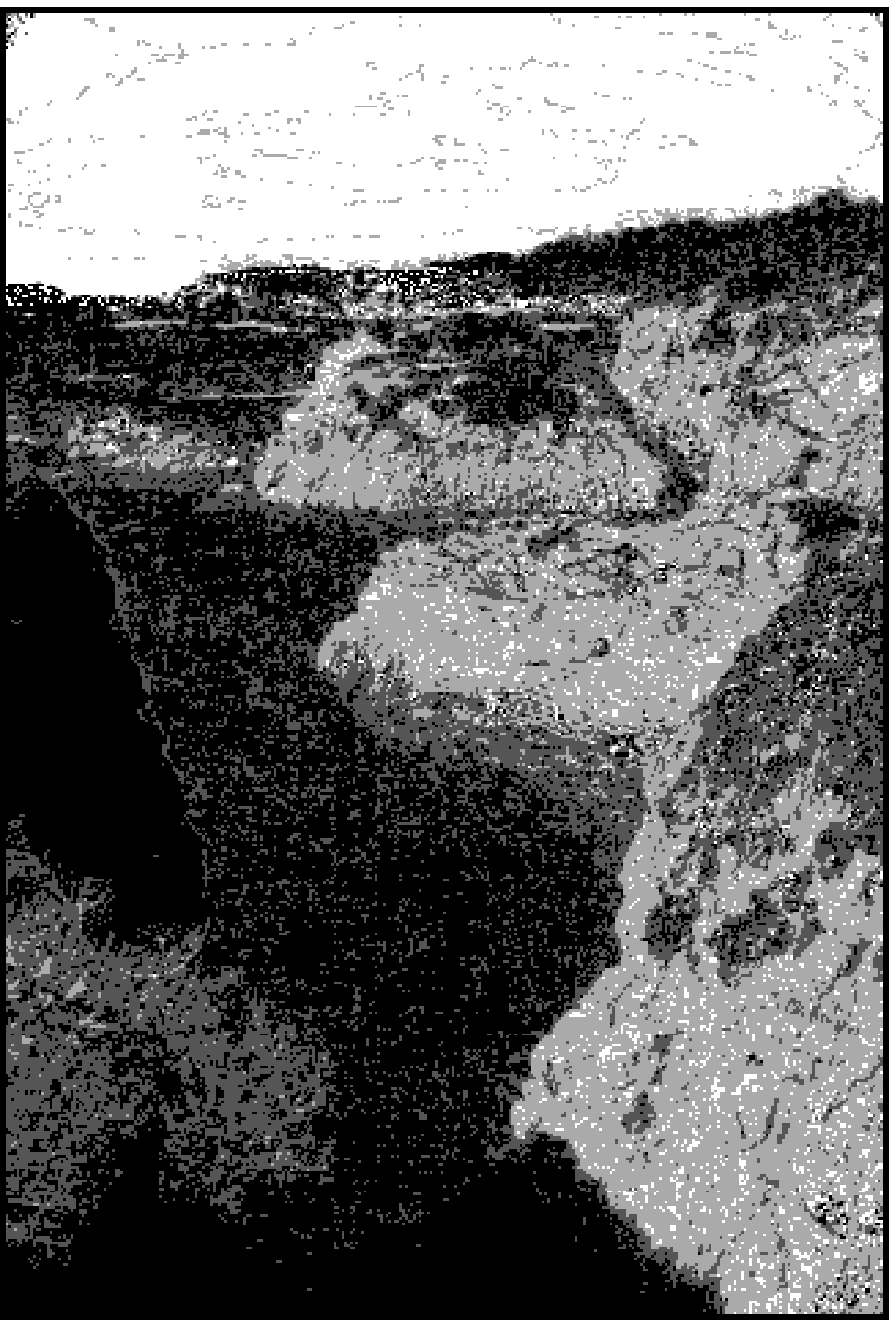}
}
\\
\subfigure[]{
\includegraphics[width=1.597in]{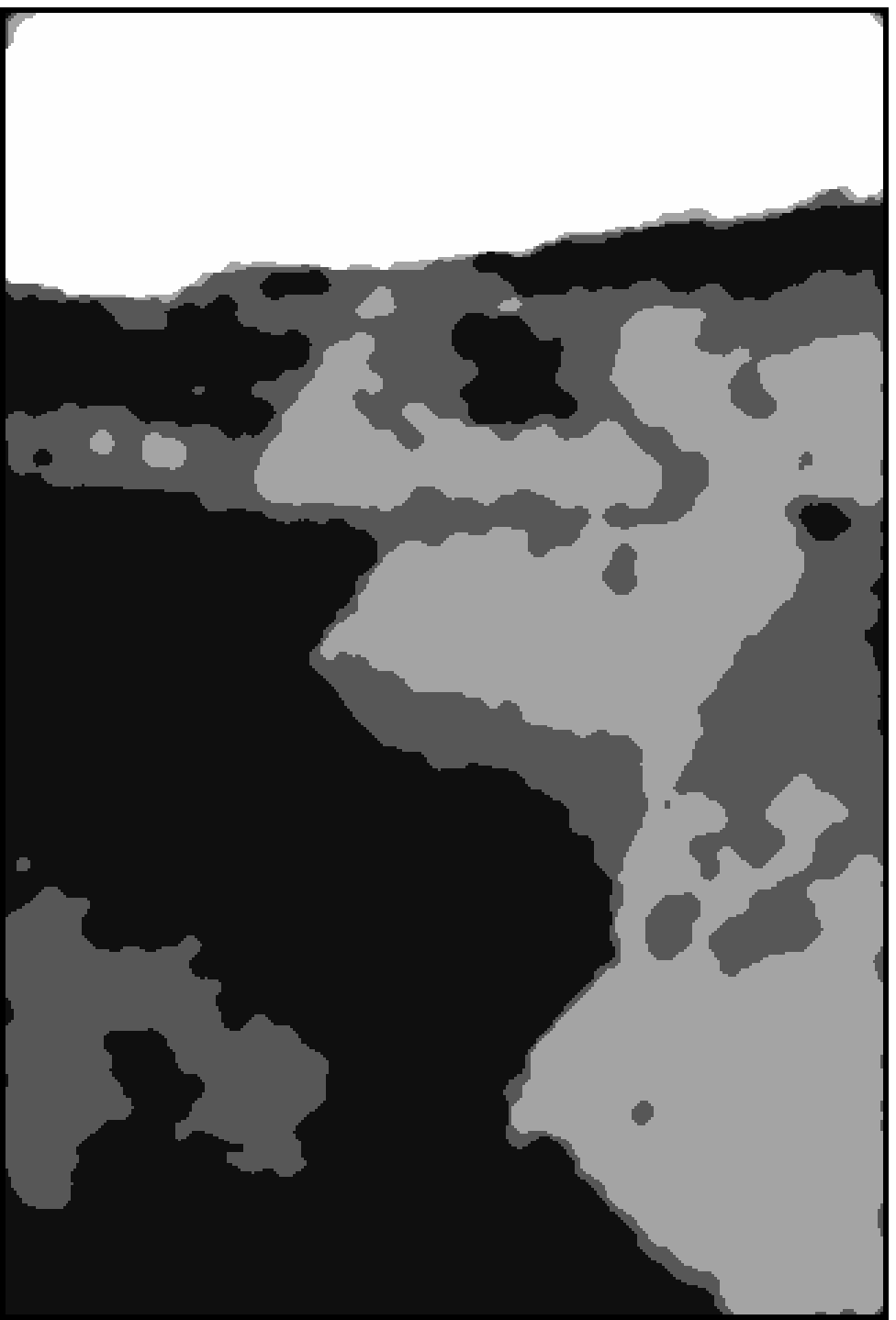}
}
\subfigure[]{
\includegraphics[width=1.597in]{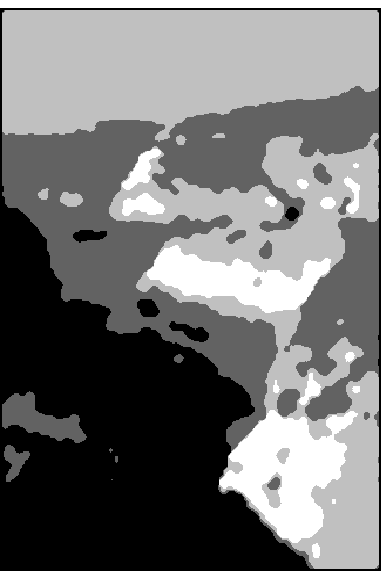}
}
\end{center}
\caption{\label{fig:photo}Grayscale photo (a) showing water, sky, vegetation, and rocks. Segmentation of image using the direct Potts method (b) with $Q=16$ and $s_t=7.5$. Image is from the Berkeley Segmentation Data Set. (See \onlinecite{martin01}.) The image at lower left (c) is a 4-level direct Potts segmentation of (a) that has been despeckled. Here the sky, rock, vegetation, and water are represented as white, light gray, dark gray, and black regions, respectively. The image at lower right (d) is the segmented and despeckled image using the four-level histogram method applied to (a).}
\end{figure}

Three examples using the direct Potts segmentation will now be discussed using a photographic image, a medical image, and an acoustic seafloor image. Each segmentation result is compared to a histogram segmentation technique in which the image is partitioned into four uniformly distributed intensity ranges. The four-level histogram method is a very efficient, linear-time process but does not distinguish grainy regions from smooth regions. The photographic image in Fig.~\ref{fig:photo}a\footnote[4]{From the Berkeley Segmentation Data Set (http://www.eecs.berkeley.edu/Research/Projects/CS/~vision/grouping/segbench/). Used with permission.} contains depictions of water, sky, vegetation, and rocks. Using step 1 of the direct Potts segmentation method as described in the previous section (with $Q=16$ and $s_t=7.5$), sky and rock are distinguished from water and vegetation according to pixel intensities. The sky and rock are represented as lighter shades, while water and vegetation are depicted as darker shades of gray in Figs.~\ref{fig:photo}b and \ref{fig:photo}c. To further separate the shades of gray, step 2 in the previous section effectively identifies regions of different texture (smooth or grainy). The full (steps one and two) direct Potts segmentation result is shown in Fig.~\ref{fig:photo}b. After filtering by using the despeckle algorithm, we see in Fig.~\ref{fig:photo}c that the sky (white, bright/smooth texture) is distinguished from the rock regions (light gray, bright/grainy), and water (dark gray, dark/smooth) is distinguished from the vegetation (black, dark/grainy). Segmenting the image by intensity using the four-level histogram produces the image in Fig.~\ref{fig:photo}d. The histogram approach does not distinguish sky from rock (both are light gray) and does comparatively less well in distinguishing vegetation (lower left of image) from the background. A comparison of Figs.~\ref{fig:photo}c and \ref{fig:photo}d underscores the importance of considering both intensity and texture in image segmentation. 

\begin{figure}
\begin{center}
\subfigure[]{
\includegraphics*[width=1.597in]{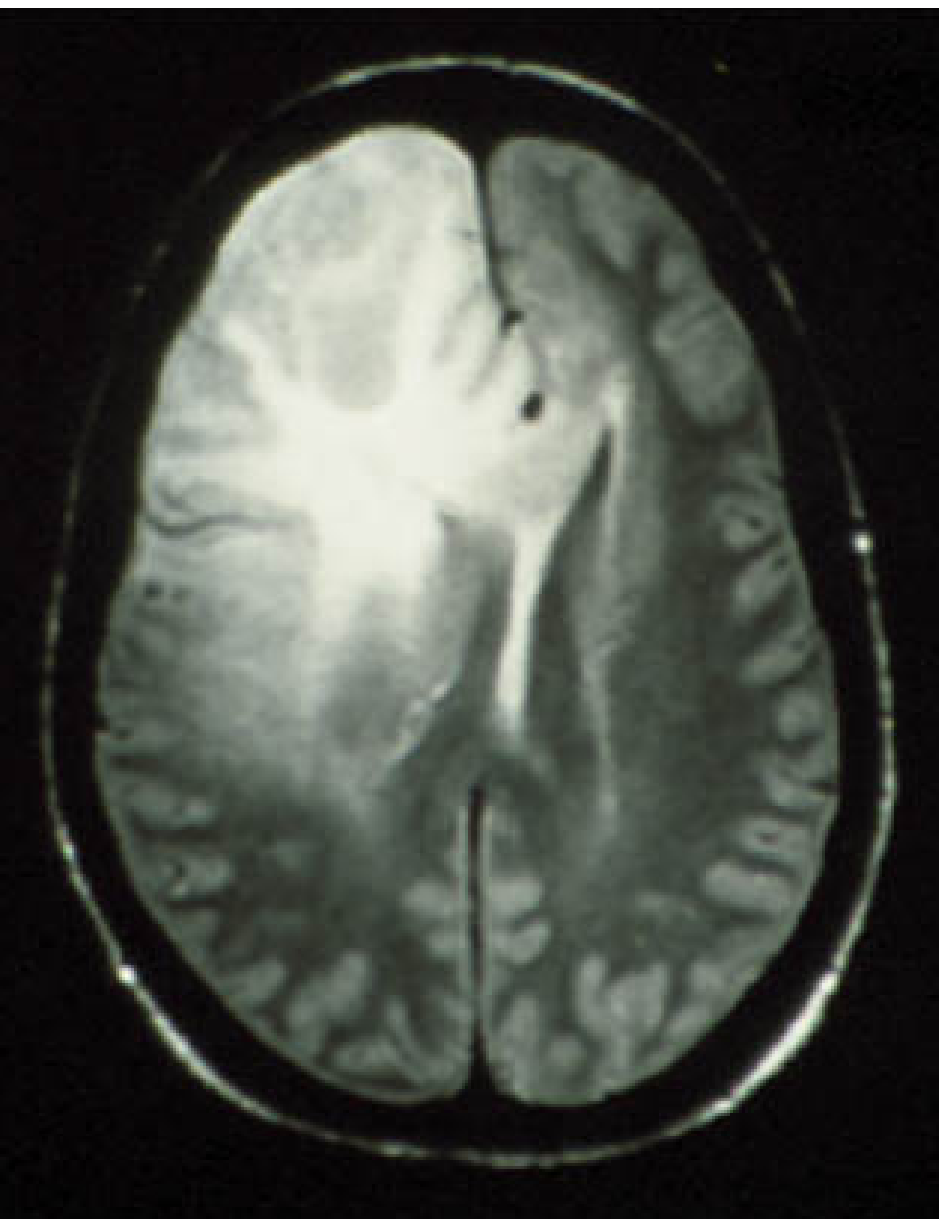}
}
\subfigure[]{
\includegraphics[width=1.597in]{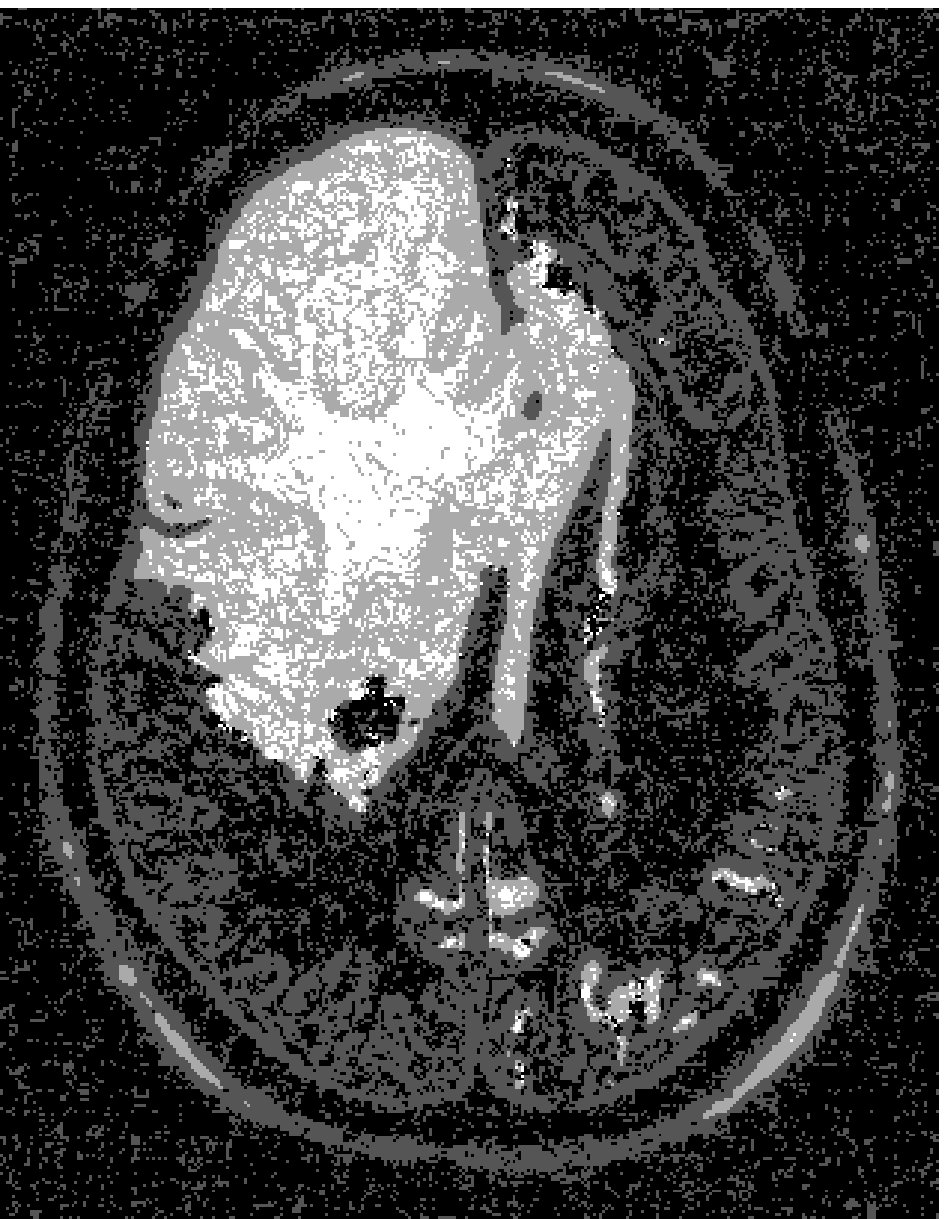}
}
\\
\subfigure[]{
\includegraphics*[width=1.597in]{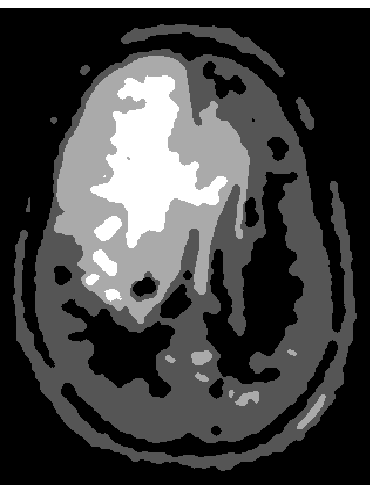}
}
\subfigure[]{
\includegraphics[width=1.597in]{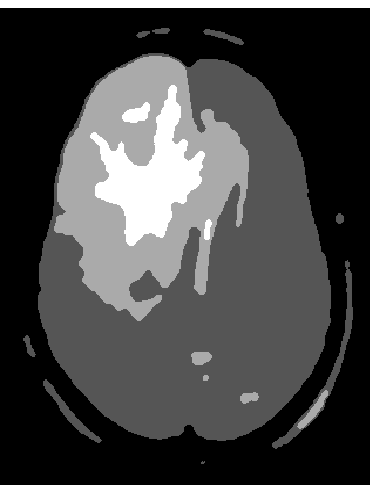}
}
\end{center}
\caption{\label{fig:mriA}MRI of a tumorous brain at upper left (a). A tumor is seen as the bright spot in the upper left region of the MRI. This image is from the Iowa Neuroradiology Library. Image at upper right (b) is the direct Potts segmentation with $Q=64$, $s_t=31.5$.  Lower left (c) is the direct Potts segmentation that has been despeckled. The clearly visible white segment identifies the tumor. A tumor is clearly visible as the bright spot in the upper left region. The periphery of the tumor is shown in light gray, while cerebrum is represented as dark gray. Other brain matter is shown in black. For comparison, lower right (d) is the despeckled four-level histogram segmentation, which identifies the tumor region, but cannot distinguish cerebrum from other brain matter.}
\end{figure}

\begin{figure}
\begin{center}
\subfigure[]{
\includegraphics[width=1.02in]{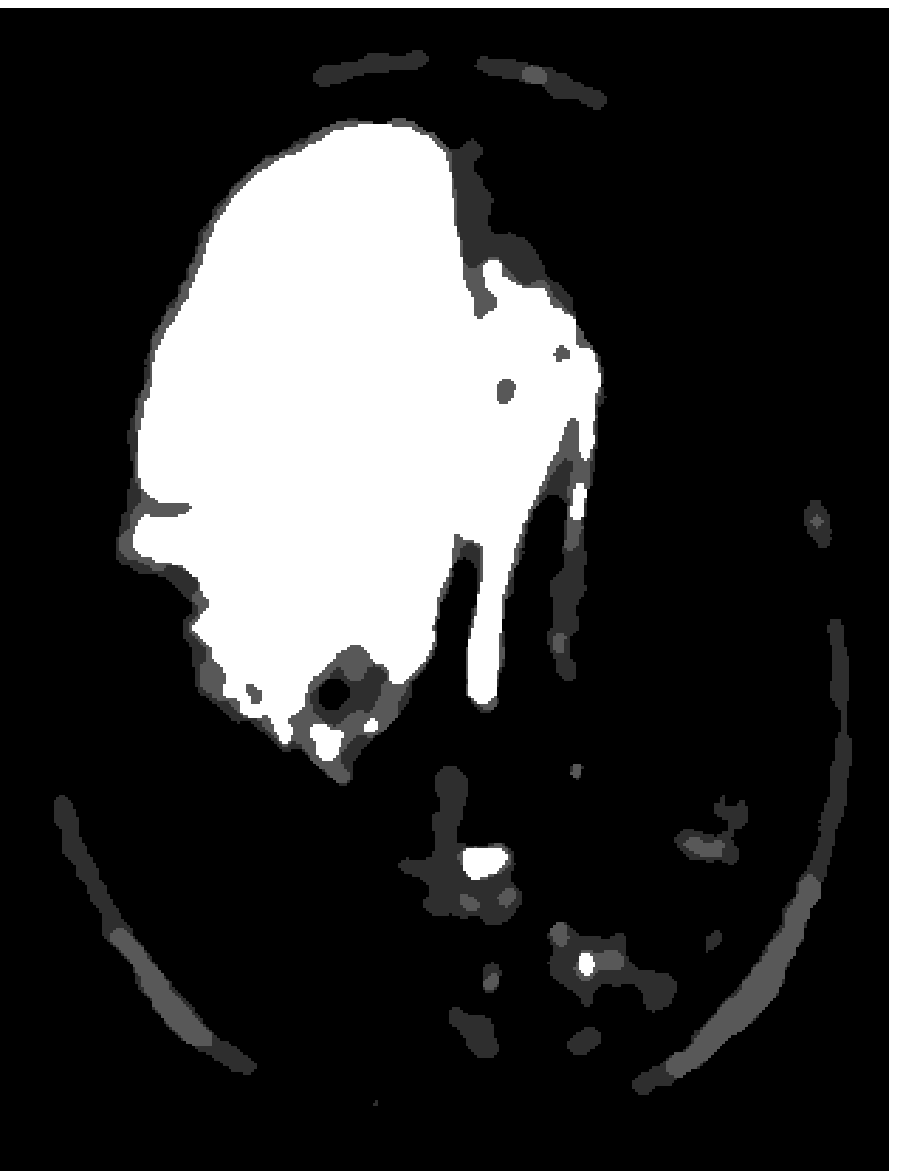}
}
\subfigure[]{
\includegraphics[width=1.02in]{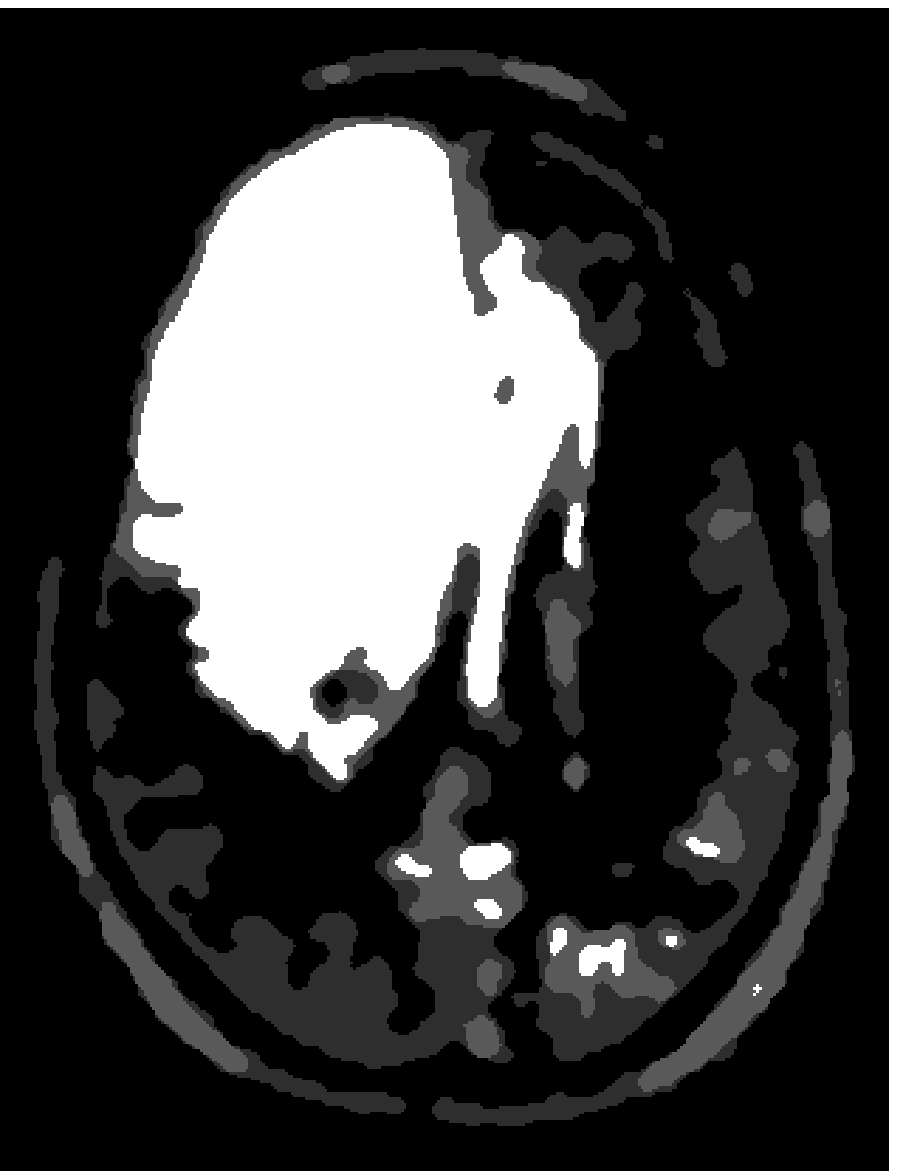}
}
\subfigure[]{
\includegraphics[width=1.02in]{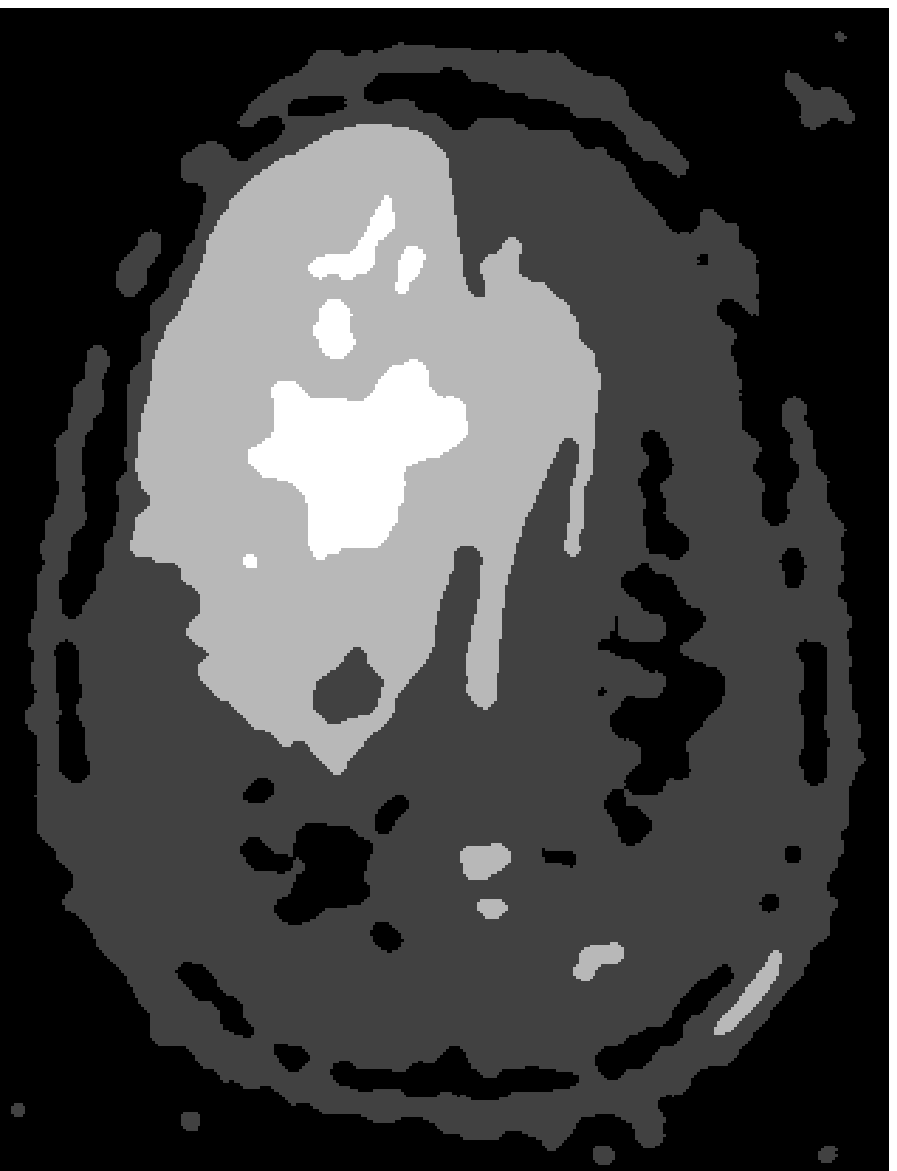}
}
\end{center}
\caption{\label{fig:brain-q}Despeckled segmentation images of Fig.~\ref{fig:mriA}a using the direct Potts method are shown for different values for $Q$. Results are shown for $Q=16$ (a), $Q=32$ (b), and $Q=128$. In each case, the intensity threshold is $s_t=(Q-1)/2$. For higher values of $Q$, more of the image is categorized as textured (light and dark gray).} 
\end{figure}

\begin{figure}
\begin{center}
\fbox{
\includegraphics*[width=3.25in]{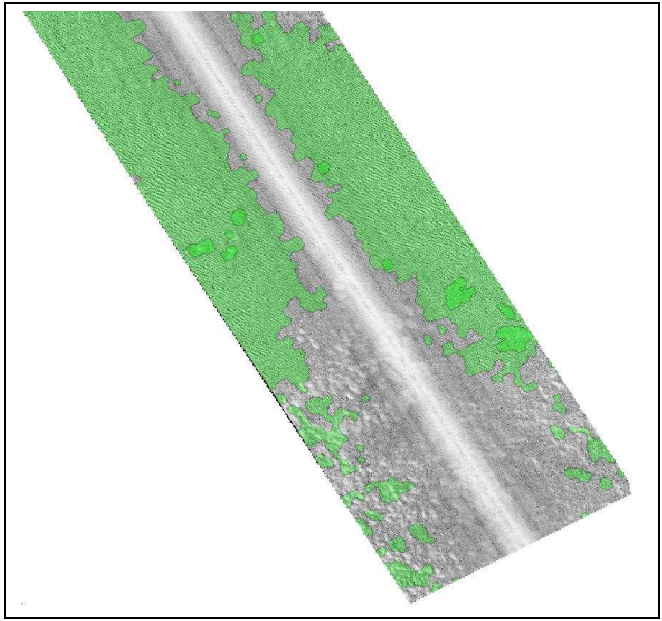}
}
\end{center}
\caption{\label{fig:seafloor}Acoustic image of the seafloor formed using a sidescan sonar system, which creates an image going out on either side via an acoustic time series. The direct Potts segmentation with $Q=16$ and $s_t=7.5$ separates the region containing sand ripples (in green) from the rest of the image. The ripples are identified as a textured region and is labeled as sand since neither mud nor rock are likely to exhibit ripples.}
\end{figure}

Figure \ref{fig:mriA}a\footnote[5]{From the Iowa Neuroradiology Library (http://www.uiowa.edu/c064s01/index.html). Used with permission.} shows an MRI image of a brain with a tumor in the upper left quadrant. The segmented image using the direct Potts segmentation method with $Q=64$ and $s_t=31.5$ is shown in Fig.~\ref{fig:mriA}b. After filtering, we obtain the the image in Fig.~\ref{fig:mriA}c. Here, the white regions may be interpreted as possible tumorous tissue, light gray corresponds to matter at the periphery of the tumor, dark gray represents cerebrum, and black is other brain matter. Automated screening for tumors may be performed by finding white regions of significant size. Although, the four-level histogram method (Fig.~\ref{fig:mriA}d) also identifies a similar tumor region, the cerebrum is not distinguished from other brain matter. The affect of the chosen value for $Q$ is demonstrated in Fig.~\ref{fig:brain-q}. Larger values of $Q$ result in more of the image being categorized as textured (represented as light and dark gray). Finally, acoustic seafloor images may be analyzed in real-time with the direct Potts segmentation method as demonstrated in Fig.~\ref{fig:seafloor}. Here sand ripples are identified (in green) as a grainy texture and is separated from the remainder of the image. Rapid seafloor characterization enables effective mine hunting/avoidance operations \cite{bentrem06,bentrem02,bentrem08}, by indicating where seafloor mines are likely to be buried (mud), partially buried (sand), or unburied (rock).

\section{Conclusions}

We have examined the direct application of the Potts model to grayscale image segmentation. The Potts model is applied to the grayscale image itself rather than its more common application to the segmented image. The result is a segmentation method which processes in linear time instead of the exponential or power-law time required with conventional Potts segmentation methods. This increase in computational efficiency is vital for real-time, high data volume, and high resolution applications. The method is analogous to minimizing the total energy in a two-dimensional magnetic spin glass where the energy is calculated according to four Potts models. Each of the four Potts models corresponds to a distinct type of magnetism, so the direct Potts segmentation method effectively distinguishes regions in the spin glass that most closely adhere to the properties of four classes of magnetism, namely ferromagnetism, antiferromagnetism, paramagnetism, and diamagnetism. In digital image segmentation, pixel intensities are substituted for the magnetic spins in the spin glass. The astute reader will note that no mention has been made of the physical basis for the coupling constants $J_\alpha$ for the regions in the segmented image. The origins of the image pixels' couplings may, in fact, be complicated or unknown. Branching of plants give a textured appearance in Fig.~\ref{fig:photo}a, while the distribution of blood vessels and other brain material likely gives the textured appearance to Fig.~\ref{fig:mriA}a. Underwater sand ripples may arise from the fluid dynamics and saltation \cite{bagnold41,kang04} at the seafloor. The direct Potts segmentation has the advantages of being unsupervised (no training sets needed), very efficient (linear-time, only a few operation per pixel), and effective for texture identification. For many image segmentation applications in which texture is important, this direct Potts method may be expected to produce better results than other linear-time methods. 

\begin{acknowledgments}
This work was supported by the Naval Research Laboratory's Advanced Graduate Research Program. The author thanks the Tulane Center for Polymer Reaction Monitoring and Characterization (PolyRMC) for resources provided. A special thanks goes to Dr. Luca Celardo for helpful discussions. The views expressed in this article are those of the author and do not represent opinion or policy of the US Navy or Department of Defense.
\end{acknowledgments}




\end{document}